# A simple setup for neutron tomography at the Portuguese Nuclear Research Reactor

M. A. Stanojev Pereira [a,b,*], J.G. Marques[a,b], R. Pugliesi [c]

[a] Instituto Tecnológico e Nuclear, Instituto Superior Técnico, Universidade Técnica de Lisboa, Estrada Nacional 10, 2686-953 Sacavém, Portugal

[b] Centro de Física Nuclear da Universidade de Lisboa, Av. Prof. Gama Pinto 2, 1699-003 Lisboa, Portugal

[c] Instituto de Pesquisas Energéticas e Nucleares, Centro do Reator de Pesquisas, Av. Lineu Prestes 2242, Cidade Universitária, 05508-000 São Paulo, Brasil

**Abstract**

A simple setup for neutron radiography and tomography was recently installed at the Portuguese Research Reactor. The objective of this work was to determine the operational characteristics of the installed setup, namely the irradiation time to obtain the best dynamic range for individual images and the spatial resolution. The performance of the equipment was demonstrated by imaging a fragment of a 17[th] century decorative tile.

# 1. Introduction

Neutron imaging techniques are important tools to investigate the internal structure of objects. The characteristics of the neutron-matter interaction makes possible to visualize hydrogen rich substances even when these are surrounded by metallic layers which would make them invisible under X-ray imaging. Thus, the neutron imaging techniques find applications in the technological fields such as the automotive, nuclear and aerospace industries, as well as in medicine, archeology, biology and geology [1,2,3,4].

A neutron image is obtained by irradiating the object in an uniform neutron beam and recording the intensity transmitted by the object. Several solutions have been used for image recording: X-ray films and track-etch foils associated to converter screens (gadolinium, dysprosium and boron), neutron scintillators coupled to Charge Coupled Devices (CCD) video cameras and neutron imaging plates [5,6,7].

The Portuguese Research Reactor (RPI) is a 1 MW open-pool type, recently converted to Low Enriched Uranium fuel [8]. Its main applications are neutron activation analysis and irradiation of electronic components with fast neutrons [9,10]. A simple setup for neutron tomography was recently installed in the thermal column of the RPI. The objective of this work was to determine the operational characteristics of the installed setup, namely, the irradiation time to obtain the best dynamic range in the images and the spatial resolution. Images obtained of a $17^{th}$ century tile fragment are shown as demonstration of its abilities. Improvements to the current setup are also discussed.

## 2. Description of the setup for neutron tomography

The tomography setup was installed at the horizontal access of the thermal column of the RPI which is radial with respect to the reactor core. The thermal column is a stacking of graphite, initially installed with over three meters of graphite, from the core edge to the beam port. This was later changed, so that presently the horizontal channel has only 90 cm of graphite until the core [8]. Fig. 1 shows a simplified view of the thermal column with its vertical and horizontal accesses. The main characteristics of the neutron beam at the irradiation position are shown in Table 1.

| Neutron flux (n.cm$^{-2}$.s$^{-1}$) | 2.2×10$^5$ |
|---|---|
| Neutron spectrum | Thermal Maxwellian (25meV) |
| n/gamma ratio (n.cm$^{-2}$.mrem$^{-1}$) | 2.1x10$^5$ |
| L/D ratio(*) | 20 |
| Cd ratio | 500 |
| Diameter (cm) | 5 |

Table 1. Characteristics of the neutron beam at the sample irradiation position.

(*) the ratio between the length and the diameter of the inlet aperture of the collimator

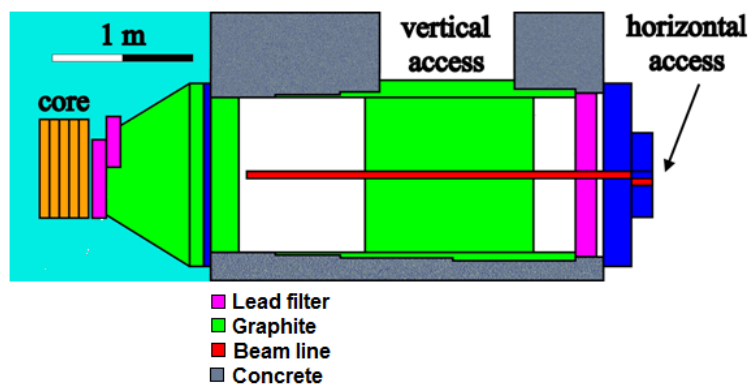

Fig. 1. Simplified view of the thermal column of the RPI with its vertical and horizontal access ports.

Fig. 2 shows a schematic diagram of the installed setup. In order to minimize gamma radiation at the irradiation position, a 10 cm lead filter is used. The image is obtained by using a neutron scintillator coupled to a CCD video camera. The scintillator screen is a 0.42 mm thick NDg type based on $^6$LiF/ZnS co-doped with Cu, Al and Au [11]. The camera is a Proline (Finger Lakes Instrumentation, USA), equipped with a Kodak KAF-1001E grade 1 CCD, with 1024 × 1024 pixels (24 × 24 μm size) which is coupled via USB to the computer [12]. Fig. 3 compiles the spectral characteristics of the NDg scintillator, its sibling ND and the CCD showing a good match between them. The lens used is a Nikon 50 mm/f1.8 and an adapter ensures a lens to CCD distance of 46.5 mm.

In order to minimize radiation damages in the CCD caused by neutrons and gammas, the camera captures the light from the scintillator through a 45⁰ mirror. It is also permanently protected by a 2.5 cm thick plate of borated polyethylene (40% boron), and a 1 cm lead plate (both with an opening for the lens). The lens is also protected by a using a 1.5 cm thick transparent Plexiglas plate, which through naked eyes creates an imperceptible optical distortion in the image. The CCD is back-illuminated, with peak efficiency of 72%.

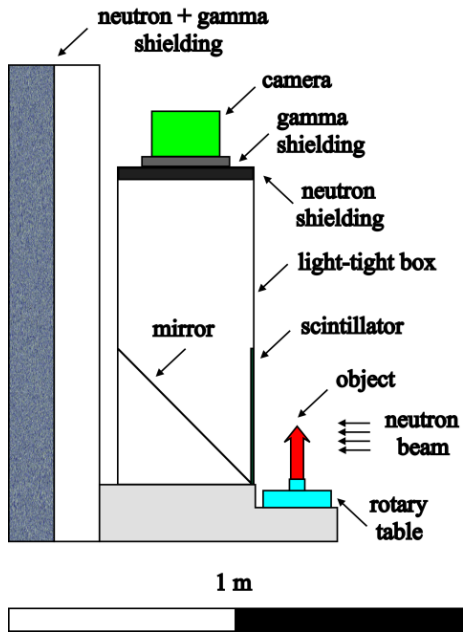

Fig. 2. Schematic diagram of the setup for neutron tomography.

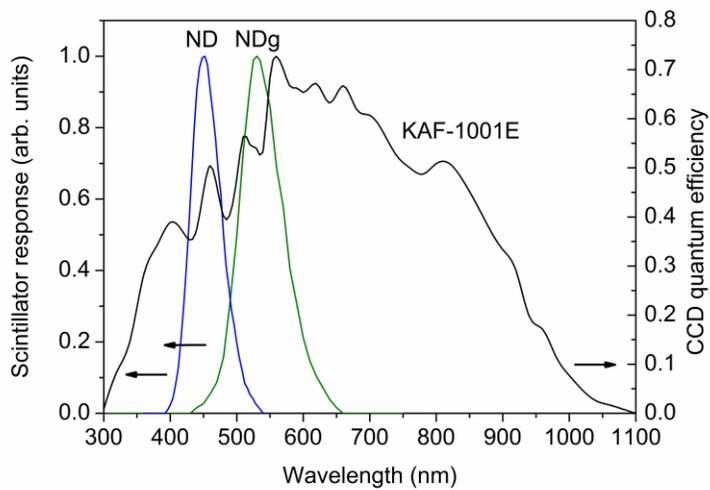

Fig. 3. Spectral characteristics of the scintillator and of the CCD.

The camera has a Peltier cooling system able to decrease the CCD temperature up to -65 ºC. The Peltier itself is cooled with chilled water (ca. 15 ºC) in a closed circuit. Fig. 4 shows the histograms for 7 minutes long dark images (no neutron beam) taken with the CCD at +21ºC and -21ºC in an 8-bit gray level (GL) scale. A strong reduction is seen on the number of pixels with significant intensity when the temperature of the CCD is lowered to -21ºC. Fig. 5 shows the variation of the number pixels with GL above the dashed line in the histograms in Fig. 4, from +21 ºC down to -21 ºC, for 7 min long dark images. The decrease in the number of 'hot' pixels follows closely the expected variation with temperature of the bulk dark current (D) in the CCD [12]:

$$D \propto T^{3/2} \exp\left(-\frac{E_G}{kT}\right) \qquad (1)$$

In eq. (1), T is the absolute temperature, k is Boltzmann's constant and $E_G$ is the energy gap energy of Si (1.11 eV at room temperature).

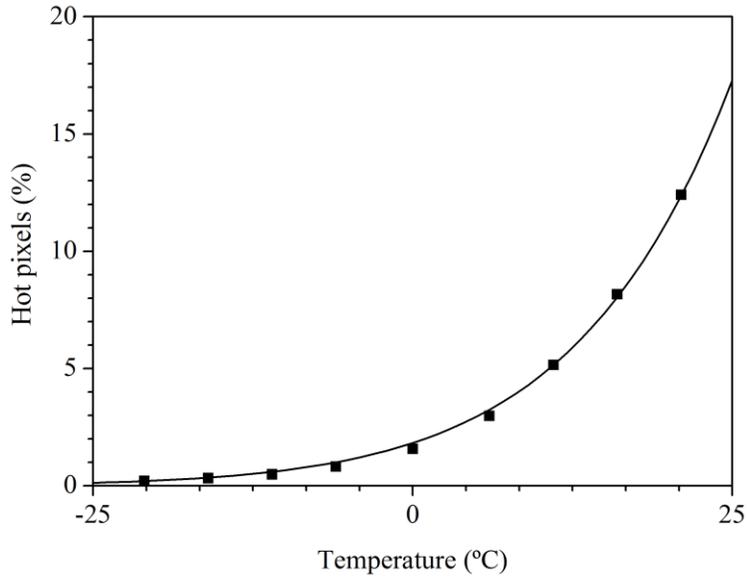

Fig. 4. Evolution of the number of hot spots in the CCD (1024 × 1024 pixels) with the temperature.

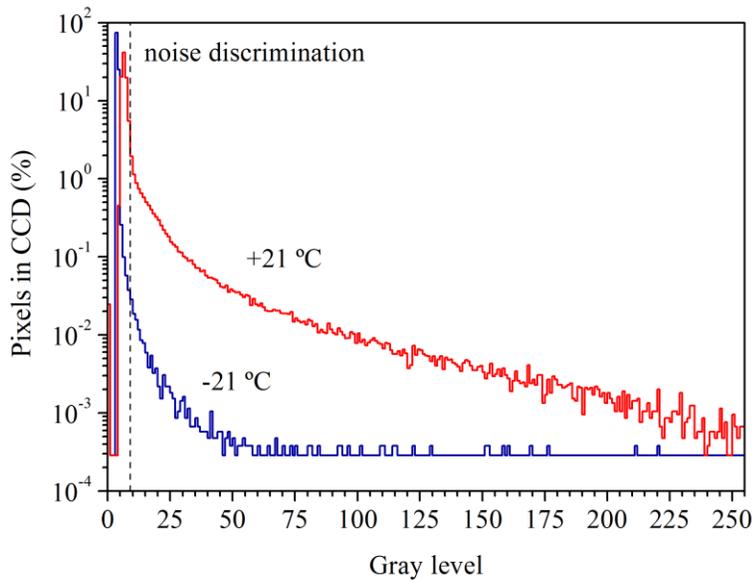

Fig. 5. Histograms of dark images taken at -21$^0$C and 21$^0$C as function of gray level.

For tomography, the object is positioned on the table (figure 2) and it is rotated after each image is acquired. The control of the camera and of the rotary table is performed through a MatLab application. The images are processed and reconstructed by using the software Octopus [13] and displayed by using the VG StudioMax 2.0 [14]. The whole system is managed by a PC with an Intel Xeon 2.4 GHz processor, with 4Gb of RAM and dual redundant hard disk drives.

**3. Data acquisition and analysis**

3.1. Dynamic range

The irradiation time to obtain the best dynamic range can be determined by means of a curve that relates GL in the image as function of the neutron exposure (E) [15]. The exposure is given by:

$$E = \phi \cdot t \tag{2}$$

where $\phi$ is the neutron flux (n/cm$^2$/s) and t is the irradiation time (s).

The fig. 6 shows the variation of "GL" as functions of "E" in the exposure interval $2.2 \times 10^5 < E < 1.3 \times 10^8$ n/cm$^2$, for the CCD at -20°C. A total of 14 points have been obtained and each one evaluated by averaging the gray level intensities of about 150,000 individual pixels in the image. The standard deviations are inserted in the points and they vary from 1% to 23% of the read value. According to the curve, the best dynamic range corresponds to a maximal exposure of $9 \times 10^7$ n/cm$^2$ or 7 minutes of irradiation and for such interval, as expected, the camera shows a linear behavior tending to saturation for higher exposures [15].

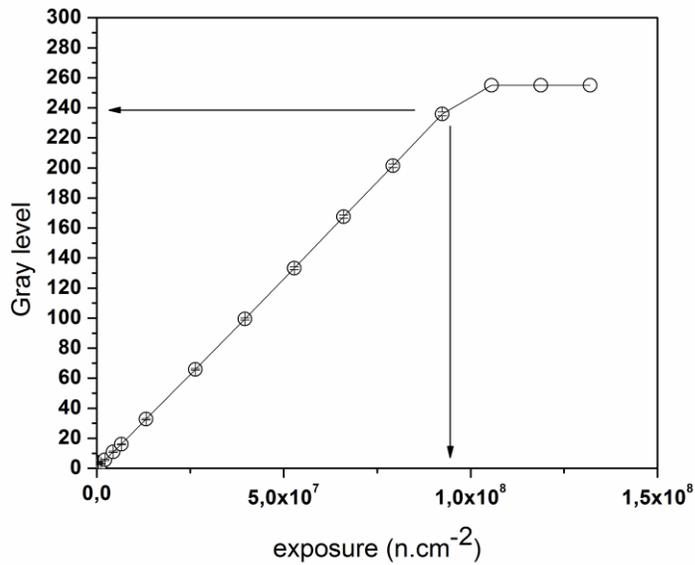

Fig. 6. Gray level as functions of exposure for the neutron tomography setup.

3.2. Spatial resolution

The spatial resolution is defined as the minimal distance between objects in such way that they can be distinguished [15]. It was determined by scanning the gray level distribution in the image of a neutron opaque edge object, here a gadolinium strip having a thickness of 127 μm which has been irradiated in a close contact to the scintillator screen. Fig. 7 shows the result obtained for 7 minutes of irradiation. The Edge Spread Function, ESF, was fitted to the points [16,17]:

$$ESF = p1 + p2 \cdot atan(p3 \cdot (x - p4)) \qquad (3)$$

where $p_1$, $p_2$, $p_3$ e $p_4$ are free parameters and x is the scanning coordinate.

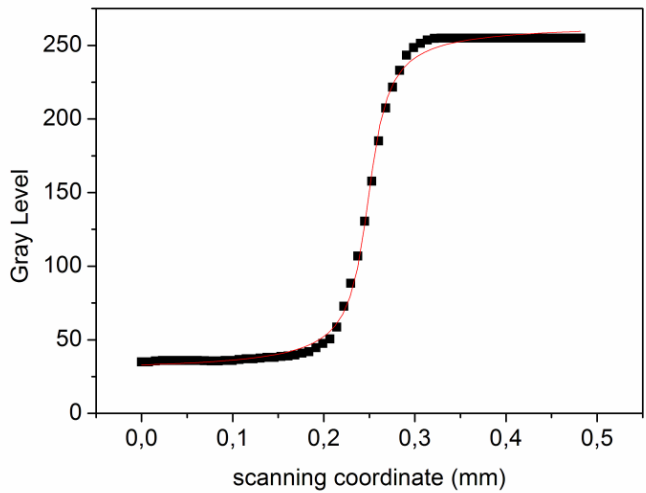

Fig. 7. Typical gray level intensity distribution and the fitted Edge Spread Function.

Usually the resolution is given by (4) and is quoted in terms of the total unsharpness "Ut" and corresponds to the width at half maximum of the Line Spread function, LSF, associated to (3) [17,18]:

$$Ut = 2/p3 \tag{4}$$

The evaluated resolution was Ut = (391 ± 10) μm, resulting of an average of ten distinct values determined in ten distinct regions of the edge object; five for the scanning performed at the horizontal direction and five for the vertical direction. Although this is the best resolution value for this equipment, this is a relatively high value, if we take into account the effective pixel size of ~200 micra (field of view of 20x20cm and the CCD with 1000x1000 pixels) and, the intrinsic resolution value for a similar camera - scintillator system [19]. The reasons for this high value can be attributed mainly to a non-perfect camera focusing, the penumbra effect since the minimal edge object - scintillator screen distance is limited to 2.5mm (the L/D ratio for the present collimator is very low (Table 1) and mainly to the Plexiglass plate inserted between the camera lens and the mirror to minimize radiation damages into the CCD (item 2).

3.3. Tomographic images

Ceramic tiles ("azulejo" in Portuguese and Spanish) are an important component of the cultural heritage in Portugal. These tiles are composed of a ceramic base covered by a vitreous glaze which gives them an impermeable glassy and brilliant surface [20]. Water can penetrate inside the tiles and contribute to their degradation. Tomographic images of a fragment of a decorative tile (Fig.8) from the 17$^{th}$ century church "Nossa Senhora dos Aflitos" located in Elvas, Portugal have been obtained and have demonstrated the potentiality of the present set up for such investigation. The images were taken with the tile dried and moistened with water. For the first condition the tile was dried at room temperature for several days and a set of 200 individual images obtained in angular steps of 0.9º have been captured. The irradiation time was 7 minutes per image and the tomography image for the angular position 0° is shown in the figure 9. As can be seen, except for some dark spots, the tile shows homogeneity in terms of neutron transmission. For the second condition, the tile was partially immersed in water during irradiation and, as before, 200 individual images have been also captured in the same previous conditions. The tomography for the angular position 0° is shown in the figure 10. The water inside the tile is visible by the darker regions of the image and the variations in water concentration are also visible by the gray level fluctuation in the image.

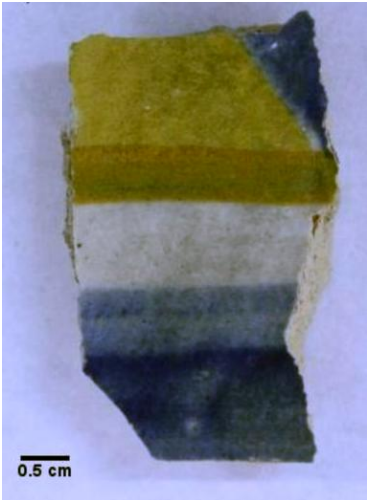

Fig. 8. Photography of the tile fragment.

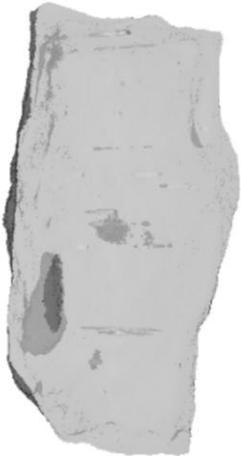

Fig. 9. Tomography of the dry tile at 0°.

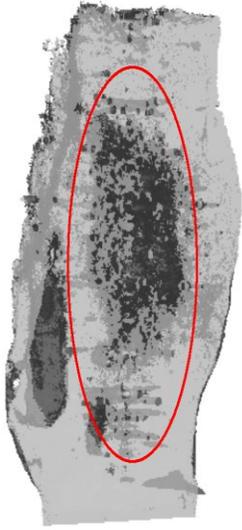

Fig. 10. Tomography of the moistened tile at 0°.

## 4. Conclusion

The obtained images have demonstrated the viability of the present neutron tomography equipment, installed at the RPI. Even for a not optimal resolution condition and for a very large irradiation time, the present image quality was enough to show important features on the water absorption by an ancient ceramic tile. In order to improve the present image quality, we are studying the possibility to install the current setup in another beam port of this same reactor, able to provide a higher thermal neutron flux ~$10^6$n/cm$^2$/s, a higher beam size of about 20cm in diameter and a better spatial resolution, less than 200µm.

### Acknowledgements

This work was partially supported by FCT, Portugal, under grant PTDC/HIS-HEC/101756 and by FAPESP, Brasil, under grant 09/50261-0. The installation for neutron tomography at the RPI was funded by FCT, Portugal, under grant

POCI/FIS/59287. The authors gratefully acknowledge Dr. Lurdes Esteves, National Tile Museum, Lisbon, for providing the tile fragment analyzed in this work.